\documentstyle[aps,preprint]{revtex}

\begin{document}

\setcounter{page}{1}
\setcounter{equation}{0}

\title{Existence of long-range order in the steady state of a \\
two dimensional, two-temperature XY model}
\vspace {1truecm}

\author{Kevin E.\ Bassler and Zolt\'an R\'acz$^*$}

\vskip 0.5truecm

\address{
Center for Stochastic Processes in Science and Engineering\\
and\\
Department of Physics\\
Virginia Polytechnic Institute and State University\\
Blacksburg, VA 24061
}
\vskip 1truecm
\maketitle

\begin{abstract}
Monte Carlo simulations are used to show that the steady state of the
$d=2$, two-temperature, diffusive XY model displays a continuous
phase transition from a homogeneous disordered phase to a phase with
long-range order.
The long-range order exists although both the dynamics and the
interactions are local, thus indicating the failure of a naive
extension of the Mermin-Wagner theorem to nonequilibrium steady states.
It is argued that the ordering is due
to effective dipole interactions generated by the nonequilibrium dynamics. \\
{}~
\end{abstract}
\narrowtext

Fluctuations increase as the spatial dimension of a system is
decreased and low-dimensional, {\em equilibrium} systems with
finite-range interactions cannot maintain long-range order.
This observation is known as the Landau-Peierls (LP) argument \cite{LP}
or, for systems with continuous symmetry,
as the Mermin-Wagner (MW) theorem
\cite{MW}. Since phase transitions and emerging long-range order
in {\em nonequilibrium} steady states
have attracted a lot of interest recently \cite{{review},{Cross}},
it is natural to ask
if similar arguments are valid for nonequilibrium systems.

Clearly, the LP argument or the MW theorem must be
generalized if we want to apply them to a nonequilibrium steady state.
The reason is that while interactions and temperature define
an equilibrium state, a nonequilibrium steady state is the result of
the interplay of interactions and dynamics. Thus apart from
restricting the range of interactions,
we must also qualify the dynamics in a nonequilibrium process.
To understand this point, consider
a kinetic Ising model in which two dynamical processes
compete with each other \cite{flipex}. The elementary processes
are (i) spin flips generated by nearest-neighbor interactions
coupled to a heat bath at temperature $T$ and (ii) random spin exchanges
generated by a $T=\infty$ heat bath. In this model, the interactions
are short ranged but the dynamics is
long-ranged because spins are exchanged at arbitrarily large distances.
As a result of long-range dynamics, effective long-range interactions
are generated and long-range order appears in this flip-and-exchange
kinetic Ising model\cite{DRT} even in $d=1$. This result is in
contradiction with the LP argument which states that long-range
order is impossible for a $d=1$ equilibrium Ising system with short-range
interactions. Thus a simple extention of the LP argument and of
the MW theorem to nonequilibrium
systems without restricting the dynamics clearly doesn't work.

One might conjecture that by restricting both the
interactions and the elementary dynamical
processes to be local, the generation of effective long-range interactions
would be prevented and an extention of the LP or MW arguments to
nonequilibrium systems could be made. This conjecture, however, is
not necessarily true since
local interactions combined with anisotropic local dynamics have been
shown to lead to effective long-range (dipole) interactions in a
two-temperature kinetic Ising model \cite{Schm}. Furthermore,
a spherical approximation to the above model has shown \cite{BZ} that
effective long-range interactions are generated not only in the
Ising case but also in the case of infinite-component order-parameter
field. Thus one may expect that neither the LP argument
nor the MW theorem can be extended to nonequilibrium steady states
even if both the interactions and the dynamics are local.
Of course, this expectation should be checked carefully since
the details of the dynamics do play an important role in determining
the ordering properties of nonequilibrium steady states, and the details,
such as the presence or absence of topological defects, the type of
defects determining the long-time relaxation of order, are indeed very
different in the Ising model, in the spherical limit and in the case of an
n-component order parameter field with $1<n\le d$.

Our aim here is to confirm the failure of the naive extention of the
MW argument by studying the existence of
long-range order in the $d=2$, two-temperature, diffusive kinetic XY model.
According to the MW theorem, long-range order cannot exist in the
$d=2$ equilibrium XY model with short range interactions. However,
we will present the results of Monte Carlo simulations which show that
a continuous transition from
the homogeneous disordered state to a state with long-range order
exists in the nonequilibrium two-temperature model.
We will also construct a field theoretic description of the transition
that is consistent with the Monte Carlo results. The
theory indicates that, just as in the two temperature
Ising case \cite{Schm}, the
critical properties are in the universality class of an equilibrium system
with long-range interactions, supporting our conjecture
that the generation of effective long
range interactions is the mechanism responsible for the breakdown of
the MW theorem.

The model we study consists of
XY spins $\vec{m}_i$, (i.\ e.\ of two dimensional vectors of unit
magnitude) at sites $i$ of a square lattice with periodic
boundary conditions imposed. There are ferromagnetic interactions of
strength $J$ between nearest-neighbor
spins, so the energy of a configuration is
$H = - J \sum_{\langle i j \rangle} \vec{m}_i \cdot \vec{m}_j$, where
$\langle i j \rangle$ indicates nearest-neighbor pairs. The system
evolves by Kawasaki exchanges \cite{Kawa} of nearest-neighbor spins,
occuring with Metropolis rates \cite{Metr}.
Exchanges along one axis of the lattice, which we call the `parallel'
direction satisfy detailed balance at an inverse
temperature $\beta_{\|} = J/T_{\|}$, while exchanges in the other
direction,
which we call the `perpendicular' direction satisfy detailed balance
at an inverse temperature $\beta_{\perp} = J/T_{\perp}$.
When the temperatures of the two heat baths are equal, the dynamics
satisfies detailed balance with the nearest neighbor XY Hamiltonian,
and the model reduces to the equilibrium kinetic XY model with Kawasaki
dynamics. However, when the two temperatures are not equal,
an energy current flows from the high-$T$ heat
bath to the low-$T$ heat bath, and detailed balance is not satisfied.
The critical behavior of the system when the two temperatures are not
equal is very interesting in the Ising version of the above model
\cite{Schm}. Its universality class coincides with that of another
equilibrium model: The Ising ferromagnet with dipolar interactions \cite{Schm}.
As we shall see below analogous situation develops in case of the XY model.

For simplicity, in our simulations, we considered
only the case $\beta_{\|} = 0$, i.e. exchanges in the
parallel direction were random. Furthermore, the distribution of
the angles of the spins (which is conserved by the dynamics) was
chosen to be uniform: for a system with $N = L_{\perp} \cdot L_{\|}$
lattice sites the angles of
the $N$ spins $\{ \vec{m}_j \}$ were $2\pi j/N$ with $j=1,2,... ,N$.
A difficulty in studying the phase transition in this system is that the
spatial anisotropy requires an analysis using anisotropic finite-size
scaling \cite{Leung}. That is, one must compare systems whose
shapes scale such that $L_{\perp}/ L_{\|}^{1+\Delta}$ is constant, where
$\Delta$ is an anisotropy exponent. As was done for the two-temperature Ising
system \cite{Danes}, and
based on renormalization group results (discussed below) which
indicate that $\Delta \approx 1$ in the present model, we choose
to simulate systems with sizes $L_{\perp} \times L_{\|} = 12
\times 9$, $16 \times 16$, $24 \times 36$, and $32 \times 64$, which are
related by the naive scaling $4L \times L^2$.

The simulations revealed a homogeneous disordered state
at small $\beta_{\perp}$, and an ordered state with long-range order at
large $\beta_{\perp}$. A typical ordered configuration is shown in Fig.~1.
Because the spin distribution is conserved by the dynamics,
the ordering occurs as a phase separation resulting in a steady-state
configuration with a spin wave in the perpendicular direction.
To study the transition with the type of ordering shown in Fig.~1, we defined
the order parameter, $\Psi$, as the following average of long-wavelength
limits of structure factors
$$
\Psi = {1 \over 2}[C_1(2\pi/L_{\perp}, 0) + C_2(2\pi/L_{\perp}, 0)]
$$
where $C_{\alpha}(q_{\perp},q_{\|})$ is the normalized Fourier transform
of the $\alpha$th component of the magnetization density. In the
simulations, we measured the time-evolution of $\Psi$ and, after
producing a rough
estimate of their relaxation times, determined the time averages
$\langle \Psi \rangle$ and $\langle \Psi^2 \rangle$ in the steady state.
The runs typically ranged
in length from $4 \times 10^5$ Monte Carlo sweeps (MCS) for the $12 \times 9$
systems to $4 \times 10^6$ MCS for the $32 \times 64$ systems.

Figs.~2 and 3 display the results of our simulations. The data for
$\Psi$ (Fig.~2) clearly show a continuous transition to a
phase-separated ordered state at $\beta_{\perp c} \approx 0.68$
(Note that the equilibrium Kosterlitz-Thouless transition \cite{Kost} occurs
at a lower temperature $\beta_{c} = 8/9$ \cite{XYcritpt}).
The presence of a continuous transition is further supported by the
cumulants for various finite-size systems (Fig.~3),
$g_L \equiv 3(1-(2/3)\langle \Psi^2
\rangle / \langle \Psi \rangle)$.
In the limit of large sizes, they cross at the
critical point $\beta_{\perp c}$, as usual for continuous
transitions \cite{Binder}.

Since Figs.2 and 3 show convincingly that the system does order, we now
turn the question of how the two-temperature XY model can have an
ordered state with long-range order.
To address this issue we construct a field theoretic description of the
model that indicates that the MW theorem fails because of the
presence of effective long-range interactions
generated by the two-temperature, diffusive dynamics.
Although not rigorous, similar arguments have previously been used to
discuss the generation of long-range interactions in Ising systems
\cite{{Schm},{Birger}}, and in the spherical limit of the two-temperature
diffusive dynamics \cite{BZ}.

We start with
the ${\cal O}(2)$ version of model B of Halprin and Hohenberg \cite{HH}
which describes critical relaxation by Kawasaki dynamics towards
the equilibrium of the coarse-grained XY model:
$$
\partial_t \vec{\phi} = \lambda \nabla^2 [ (-\nabla^2 + \tau)\vec{\phi} +
{1 \over 3!} g \vec{\phi}\phi^2 ] + \vec{\eta}
$$
Here, $\vec{\phi}({\bf x},t)$ is the
two-dimensional, coarse-grained order parameter, the parameters
$\lambda$ and $g$ are constants, and the relevant temperature dependence
is contained in $\tau$, such that
the critical point corresponds to $\tau=0$.  Furthermore,
$\vec{\eta}({\bf x},t)$ is a Gaussian noise source which has zero mean and
$ \langle \eta_{\alpha}({\bf x},t)
\eta_{\beta}({\bf x}',t') \rangle =
2 \lambda \delta_{\alpha \beta} \delta (t-t')\nabla^2
\delta (\bf {x}-\bf {x}')$.

Now consider the generalization of model B to a $d$-dimensional
two-temperature model with spin exchanges
that occur at one temperature in an $m$-dimensional `parallel' subspace
and at another temperature in the remaining $d-m$ `perpendicular' dimensions.
In order to account for the different temperatures of the exchanges
in the different subspaces of this model,
the Laplacian operators and the noise term must be split into parallel
and perpendicular parts. Furthermore, parameters such as
$\tau$ and $g$ will have different values
depending whether they are associated with the diffusion in the
parallel or the perpendicular directions. For example,
the model B term $\tau \nabla^2 \vec{\phi}$ will split into two terms:
$\tau_\perp \nabla_\perp^2 \vec{\phi}$ and $\tau_{\|} \partial^2
\vec{\phi}$, where $\partial$ ($\nabla_\perp$) indicates a gradient
over the $m$ parallel ($d-m$ perpendicular) directions.
As $\tau$ has now been split into parallel and perpendicular parts,
the theory can describe various types of critical behavior depending
on whether: (i) $\tau_\perp=0$ and $\tau_\| > 0$, (ii) $\tau_\| =0$ and
$\tau_\perp > 0$, or (iii) $\tau_\perp= \tau_\|=0$. The type of ordering
ordering seen in Fig.1 corresponds to (i)
and in the following we restrict our consideration to that case.
Keeping only those terms which are relevant
(in the renormalization group sense)
near the upper critical dimension $d_c = 4 - m$,
we arrive at the following equation
$$
\partial_t \vec{\phi} = \lambda \nabla^2_\perp [ (-\nabla^2_\perp + \tau_{\|}
{\partial^2 \over \nabla^2_\perp} + \tau_{\perp})\vec{\phi}+ {1 \over 3!}
g_{\perp} \vec{\phi}\phi^2 ] + \vec{\eta}_{\perp} .
$$
Note that only the perpendicular part of the noise,
$\vec{\eta}_{\perp}$, is relevant near $d_c$.
It has vanishing mean and
$ \langle \eta_{\perp \alpha}({\bf x},t)
\eta_{\perp \beta}({\bf x}',t') \rangle =
2 \lambda \delta_{\alpha \beta} \delta (t-t')\nabla^2_\perp
\delta (\bf {x}-\bf {x}')$.

Interestingly, this Langevin equation is of the form
$$
\partial_t \vec{\phi} = \lambda \nabla^2 {\delta {\cal H} \over \delta
\vec{\phi}} + \vec{\eta}
$$
where $\vec{\eta}$ satisfies the fluctuation dissipation theorem
and therefore the above equation describes the critical dynamics of an
equilibrium system.
The Hamiltonian, ${\cal H}$, of that equilibrium system is
most easily expressed in Fourier space as
$$
{\cal H} = {1 \over 2}\int_{k} {k_{\perp}^4 + \tau_{\|}k_{\|}^2  +
\tau_{\perp}k_{\perp}^2 \over k_{\perp}^2} \vec{\phi}(k) \vec{\phi}(-k)
+ {g_{\perp} \over 4!} \int_{k_1, \ldots, k_4} \vec{\phi}(k_1)\vec{\phi}
(k_2)\vec{\phi}(k_3)\vec{\phi}(k_4)\delta(k_1+\ldots + k_4) ,
$$
which is the Hamiltonian of the the XY model with dipolar interactions.
Thus, as in the Ising case \cite{Schm},
we arrive at the conclusion that the mesoscopic critical properties of the
two-temperature diffusive XY model
are in the same universality class as an equilibrium model
with long-range (dipole) interactions.
Most importantly for the two-temperature XY model, however, we see
the mechanism that apparently causes the Mermin-Wagner theorem to fail:
the two-temperature diffusive dynamics generate
effective long-range interactions in the steady state of the system.

The static critical properties of the equilibrium XY model with dipole
interactions have been studied using renormalization group methods for
the $m=1$ case \cite{Brez}. The results
show that the static structure function scales as
$$
S(\vec{k}_{\perp}, \vec{k}_{\|},\tau)
= \mu^{-2+\eta} S(\vec{k}_{\perp}/\mu, \vec{k}_{\|}/\mu^{1+\Delta},\tau/
\mu^{1/\nu})
$$
which defines the exponents $\eta$, $\nu$, and $\Delta$. To second order
in $\varepsilon = 3 - d$, it has been found that
$
\nu \approx 0.500 + 0.100 \varepsilon + 0.054 \varepsilon^2$,
$\eta \approx 0.0177 \varepsilon^2$, and
$\Delta = 1 - \eta /2$.

Fig.~4 shows the scaling plot of $\langle \Psi \rangle$ using the above
values of the exponents with $\varepsilon = 1$. The data collapse is not
perfect, but it is consistent with the predictions of the field theory,
considering that the field-theory
predictions are only approximate, being to finite order in $\varepsilon$.
We caution, however, that a range of exponents ($\nu =0.6\pm 0.1$ and
$\eta =0.10\pm 0.15$) also produce a collapse of
the data. A more accurate determination of the value of the
exponents would require data for a variety of system shapes and
significantly larger system sizes than were used in the current study.

In summary, our MC simulations taken together with the field-theoretic
results suggests a simple picture: Long-range order does exist in the
$d=2$, two-temperature, diffusive XY model and it is caused by effective
long-range interactions generated by the anisotropic
diffusive dynamics coupled with the violation of the fluctuation-dissipation
theorem.  In light of this result, we return to the question posed
at the outset: Can the LP argument or the MW theorem be extended to
nonequilibrium steady states? In general, the answer to this question
appears to be no, even for systems with
purely local interactions and dynamics.  However, it still remains
possible that the LP arguement or the MW theorem can
be extended to nonequilibrium steady-state systems which evolve
with some restricted class of dynamical rules.
Unfortunatly, our knowledge in this field is rather limited; we know
only that long-range order can appear in low-dimensional
nonequilibrium systems whose dynamics is either long-ranged\cite{DRT}, or
involves anisotropic diffusion\cite{review}.
Clearly, much more work needs to be done to fully answer the question posed
in this paper.

\bigskip

\noindent{\bf Acknowledgements:}
We thank N. Menyhard, B. Schmittmann and R.K.P. Zia for useful discussions.
This work was partly supported by the National Science Foundation
through the Division of Materials Research.

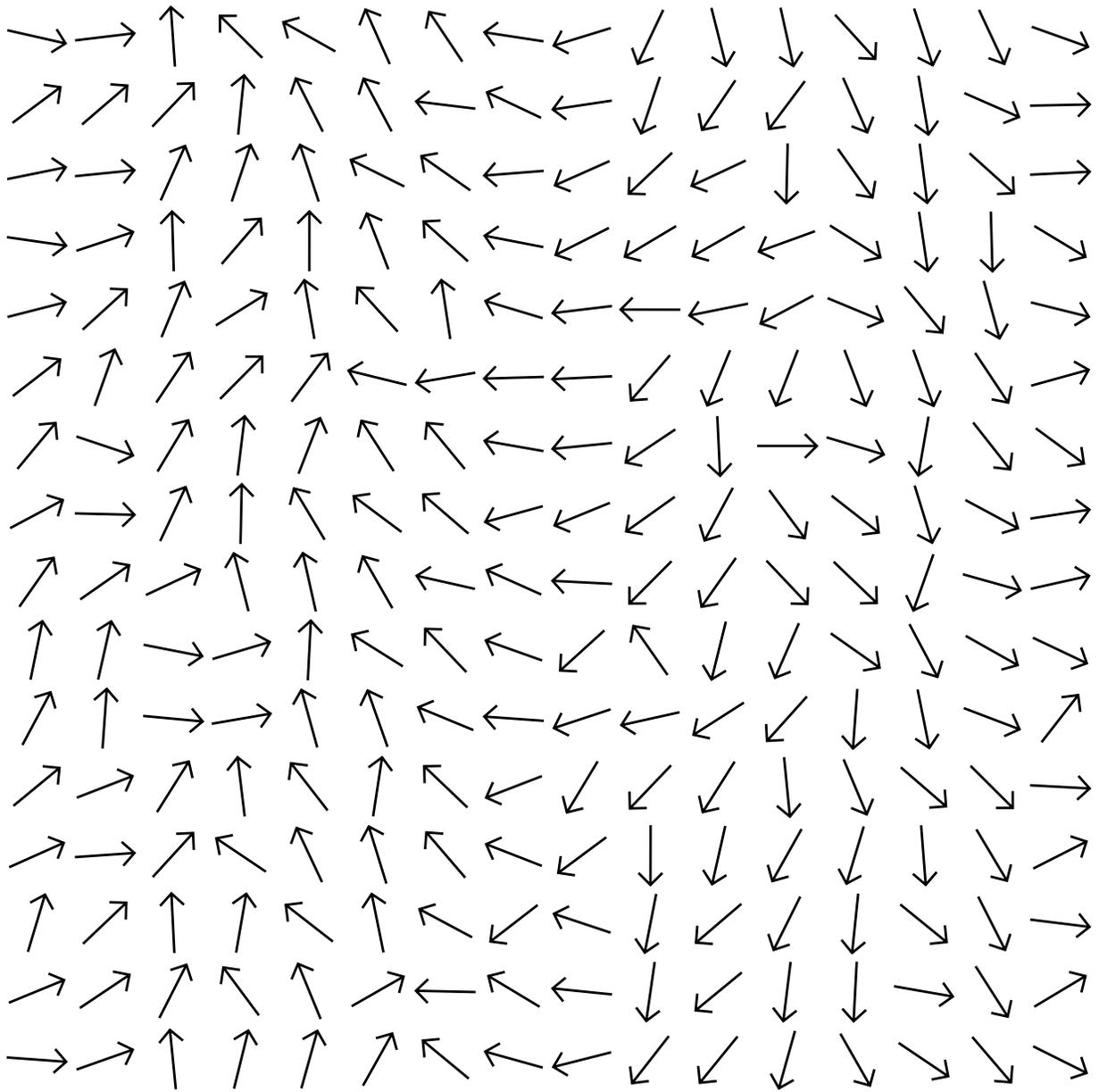
\begin{figure}
\caption{
A typical ordered configuration with long-range order. The $\perp$
and $\|$ directions correspond to the directions of the $\perp$
and $\|$ bonds, respectively. The configuration shows a spin wave
in the perpendicular direction.
}

\end{figure}

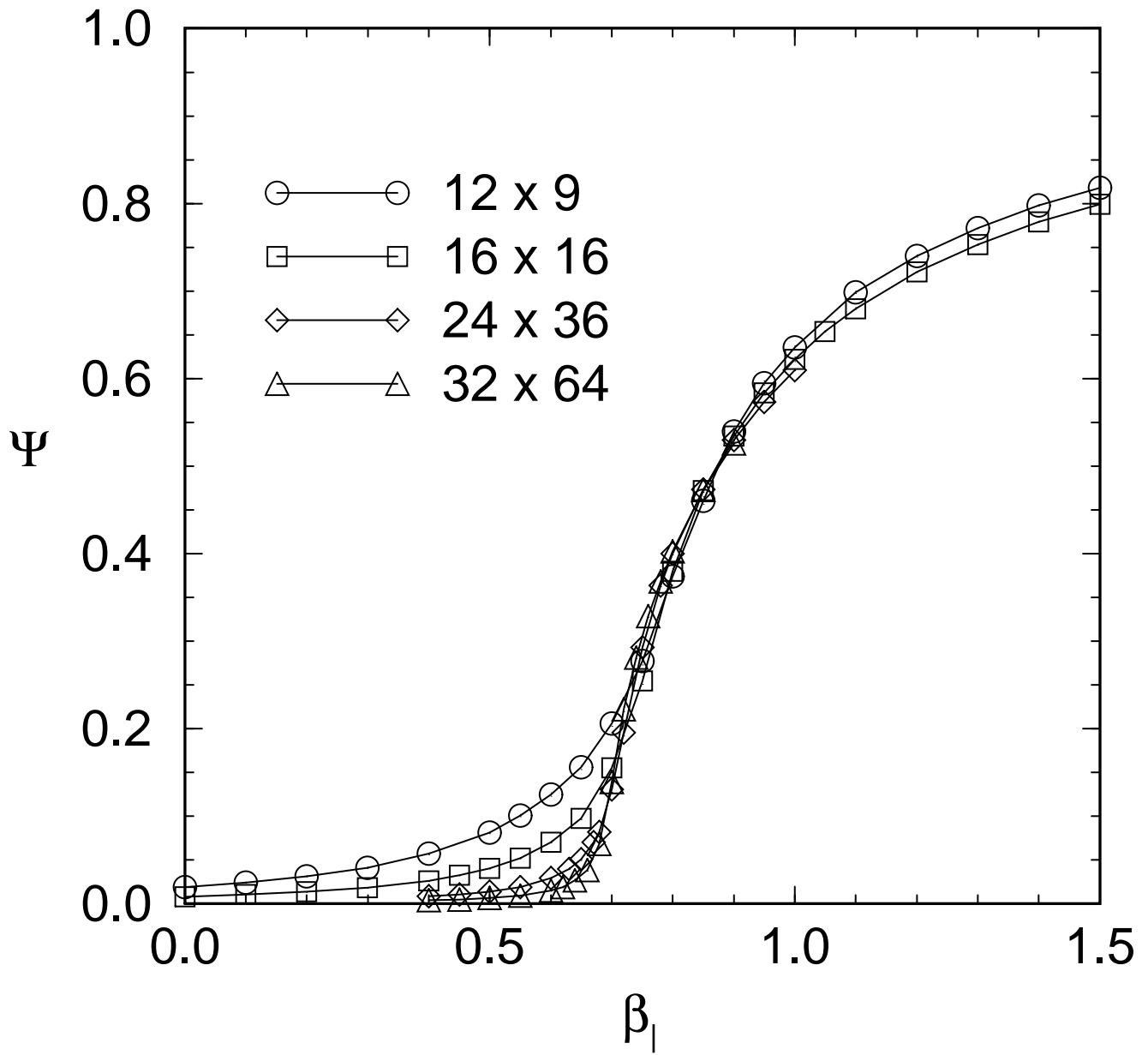
\begin{figure}
\caption{
Monte Carlo results for the order parameter $\langle \Psi \rangle$,
showing a transition to long-range order at $\beta_{\perp c} \approx 0.68$.
The size and shape ($L_{\perp} \times L_{\|}$) of the systems is indicated
in the legend.
Error bars are much smaller than the symbol size.
}

\end{figure}

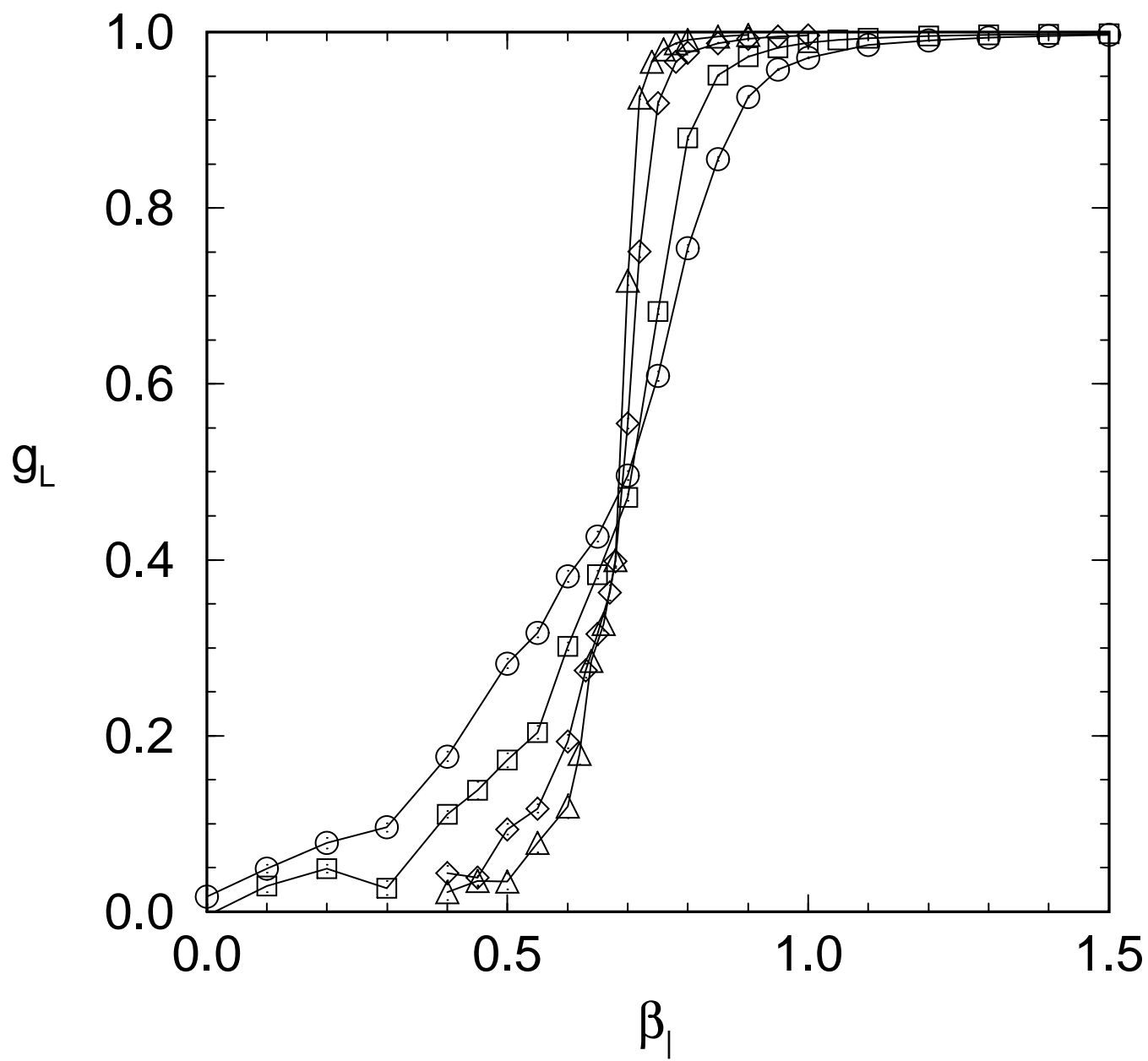
\begin{figure}
\caption{
Monte Carlo results for the cumulant $g_L$ defined in the text. The
data for the different finite-size systems cross asymptotically at
$\beta_{\perp c} \approx 0.68$, in agreement with the data for
$\langle \Psi \rangle$, shown in Fig.~2.
The size and shape, $L_{\perp} \times L_{\|}$, of the systems is the
same in Fig.~2.
Error bars indicate one standard deviation of statistical errors.
}

\end{figure}

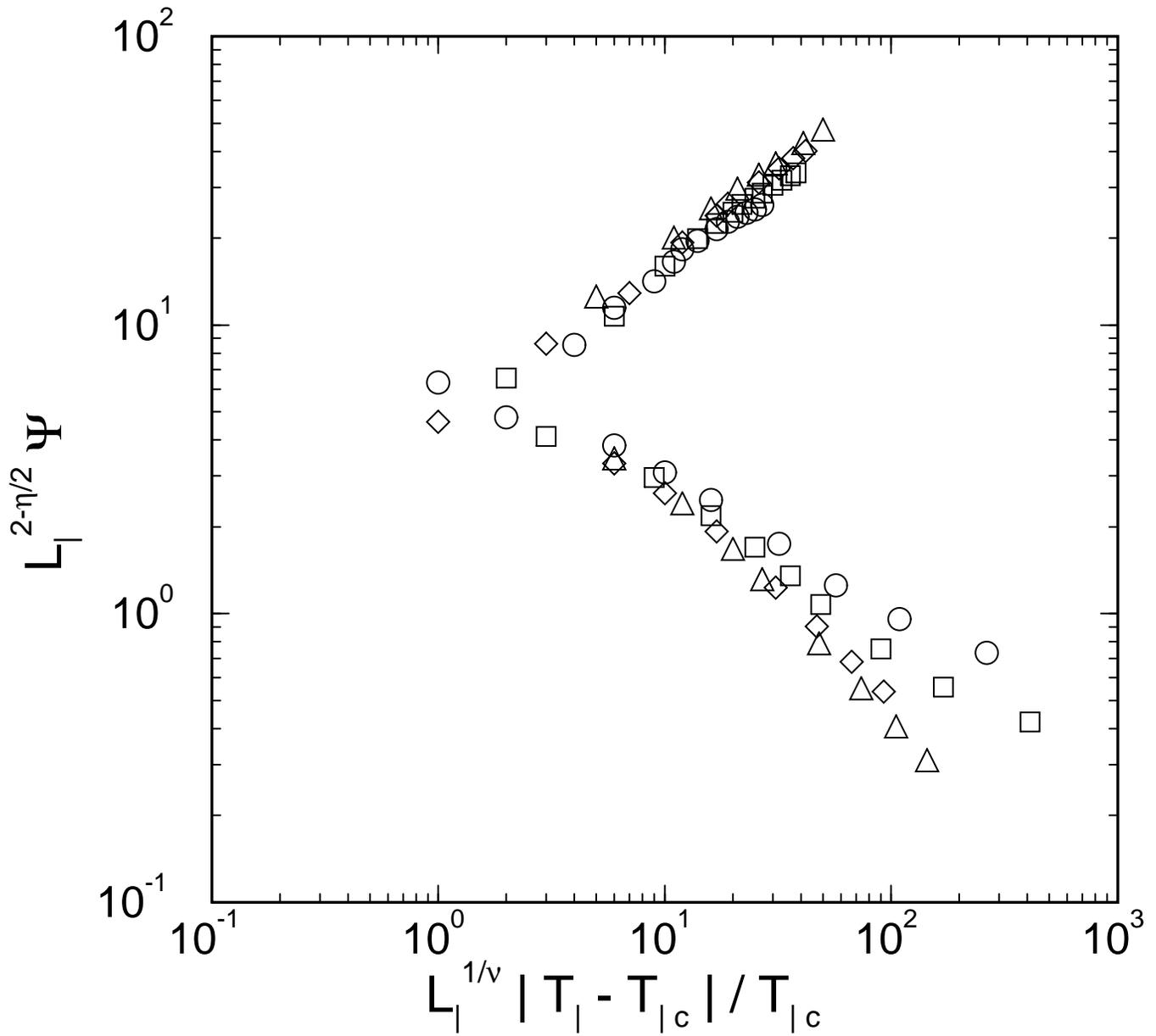
\begin{figure}
\caption{
Finite-size scaling of Monte Carlo results for the order parameter
$\langle \Psi \rangle$, using $\beta_{\perp c}=0.68$, $\nu = 0.65$, and
$\eta = 0.02$. The size and shape, $L_{\perp} \times L_{\|}$,
of the systems is the same as in Fig.~2.
}

\end{figure}


\begin{references}
\bibitem[*]{Auth} Current and Permanent address: Institute for Theoretical
Physics, E\"otv\"os University, 1088 Budapest, Puskin u. 5-7, Hungary.

\bibitem{LP} L.D. Landau and E.M. Lifshitz, {em Statistical Mechanics}
(Pergamon, London, 1981).

\bibitem{MW} N.D. Mermin and H. Wagner, Phys.\ Rev.\ Lett.\ {\bf 17},
1133 (1966).

\bibitem{review} For a comprehensive review and further references, see:
B. Schmittmann and R.K.P. Zia, to appear in {\em Phase Transitions and
Critical Phenomena}, eds. C. Domb and J.L. Lebowitz, (Academic Press, N.Y.).

\bibitem{Cross} M.C. Cross and P.C. Hohenberg, Rev. \ Mod. \ Phys. {\bf 65},
851 (1994).

\bibitem{flipex} A. DeMasi, P.A. Ferrari, and J.L. Lebowitz,
Phys.\ Rev.\ Lett.\ {\bf 55}, 1947 (1985).

\bibitem{DRT} M. Droz, Z. R\'acz, and P. Tartaglia,
Phys.\ Rev. {\bf A41}, 6621 (1989).

\bibitem{Schm} B. Schmittmann, Europhys.\ Lett.\ {\bf 24}, 109 (1993).

\bibitem{BZ} K.E. Bassler and Z. R\'acz, Phys.\ Rev.\ Lett.\
{\bf 73}, 1320 (1994).

\bibitem{Kawa} K. Kawasaki, Phys.\ Rev.\ {\bf 145}, 224 (1963).

\bibitem{Metr} N. Metropolis, A. Rosenbluth, M. Teller, and A. Teller,
J.\ Chem.\ Phys.\ {\bf 21}, 1087 (1953).

\bibitem{Leung} K.-t. Leung,
Phys.\ Rev.\ Lett.\ {\bf 66}, 453 (1991).

\bibitem{Danes} E.L. Praestgaard, H. Larsen, and R.K.P. Zia,
Europhys.\ Lett.\ {\bf 25}, 447 (1994).

\bibitem{Kost} J.M. Kosterlitz and D.J. Thouless, J.\ Phys.\ C {\bf 6},
1181 (1973).

\bibitem{XYcritpt} See, for example, J.F. Fernandez, M.F. Ferreira,
J. Stankiewicz, Phys.\ Rev.\ {\bf B34}, 292 (1986).

\bibitem{Binder} K. Binder, Z.\ Phys.\ {\bf B43}, 119 (1981).

\bibitem{Birger} B.Bergersen and Z. R\'acz,
Phys.\ Rev.\ Lett. {\bf 67}, 3047 (1991).

\bibitem{HH} P.C. Hohenberg and B.I. Halperin, Rev.\ Mod.\ Phys.\ {\bf 49},
435 (1977).

\bibitem{Brez} E. Br\'ezin and J. Zinn-Justin, Phys.\ Rev.\ {\bf B 13},
251 (1976). Note that this paper contains typos, which we have corrected to
obtain the quoted results.

\end{references}
\end{document}